\documentclass[aps,pra,twocolumn,superscriptaddress]{revtex4-1} 

\usepackage{graphicx,subfigure}
\usepackage{amsmath,amssymb,bm}
\usepackage{mathtools}
\usepackage{multirow}
\usepackage{textcomp}
\usepackage{float}
\usepackage{xcolor}
\usepackage[normalem]{ulem}
\usepackage{dsfont}
\usepackage{url}

\newcommand{\ket}[1]{\ensuremath{\left\vert{#1}\right\rangle}}

\newcommand{\bra}[1]{\ensuremath{\left\langle{#1}\right\vert}}

\renewcommand{\vec}[1]{\ensuremath{\mathbf{#1}}}

\usepackage{mathtools}
\usepackage{tcolorbox}
\tcbset{colback=blue!5,colframe=blue!40!white}
\newcommand{\tr}[1]{\mathrm{Tr}\left[ #1 \right]}
\newcommand{\Tr}{\mathrm{Tr}}
\newcommand{\ave}[1]{\left \langle #1 \right \rangle}

\newcommand{\inp}[2]{\left \langle #1 \middle | #2 \right \rangle}

\begin{document}

\title{Calculating R\'{e}nyi Entropies with Neural Autoregressive Quantum States}
\author{Zhaoyou Wang}
\thanks{zhaoyou@stanford.edu}
\affiliation{Department of Applied Physics, Stanford University, Stanford, California 94305, USA}
\author{Emily J. Davis}
\thanks{emilyjd@stanford.edu}
\affiliation{Department of Physics, Stanford University, Stanford, California 94305, USA}

\begin{abstract}
Entanglement entropy is an essential metric for characterizing quantum many-body systems, but its numerical evaluation for neural network representations of quantum states has so far been inefficient and demonstrated only for the restricted Boltzmann machine architecture. Here, we estimate generalized R\'{e}nyi entropies of autoregressive neural quantum states with up to $N=256$ spins using quantum Monte Carlo methods. A naive ``direct sampling'' approach performs well for low-order R\'{e}nyi entropies but fails for larger orders when benchmarked on a 1D Heisenberg model. We therefore propose an improved ``conditional sampling'' method exploiting the autoregressive structure of the network ansatz, which outperforms direct sampling and facilitates calculations of higher-order R\'{e}nyi entropies in both 1D and 2D Heisenberg models. Access to higher-order R\'{e}nyi entropies allows for an approximation of the von Neumann entropy as well as extraction of the single copy entanglement. Both methods elucidate the potential of neural network quantum states in quantum Monte Carlo studies of entanglement entropy for many-body systems.
\end{abstract}
\date{\today}

\maketitle

\section{Introduction}
Quantum entanglement is a fundamental property underlying diverse phenomena in condensed matter and gravitational systems~\cite{amico2008entanglement,harlow2016jerusalem} and provides the essential resource enabling quantum information technologies~\cite{horodecki2009quantum}. Entanglement entropy quantifies the amount of entanglement across a cut in a quantum state, and can reveal emergent behavior such as topological order~\cite{kitaev2006topological,levin2006detecting,flammia2009topological} and quantum phase transitions~\cite{osborne2002entanglementb, vidal2003entanglement, calabrese2004entanglement}. The set of R\'{e}nyi entropies
\begin{equation}
    S_n(\rho_A) = \frac{1}{1-n} \ln \tr{\rho_A^n}, n \geq 0,
\end{equation}
encodes the full entanglement spectrum~\cite{calabrese2008entanglement,li2008entanglement} for $\rho_A$, the reduced density matrix for a bipartition of a pure state $\ket{\psi}\in \mathcal{H}_A \otimes \mathcal{H}_B$. The second R\'{e}nyi entropy $S_2$ is often most feasible to measure numerically~\cite{hastings2010measuring} and experimentally~\cite{kaufman2016quantum}, especially compared to the von Neumann entropy $S_1$, but a wealth of information can also be gleaned from the less-accessible R\'{e}nyi entropies at higher orders $n$. In the limit $n \rightarrow \infty$, the single copy entanglement $S_\infty$ measures the distillable maximally entangled pairs from a single copy of a quantum state~\cite{eisert2005single,dimic2018single}, and for some critical systems is directly proportional to the von Neumann entropy~\cite{orus2006half,peschel2005single}. More generally, high-order $S_n$ are dominated by the low-lying levels of $\rho_A$, which may serve as order parameters~\cite{thomale2010entanglement,de2012entanglement}. Beyond yielding insight into quantum states, different R\'{e}nyi entropies can provide information about operator spreading and thermalization via their quench dynamics~\cite{rakovszky2019sub}.

Numerical techniques have been developed to estimate entanglement entropy for quantum many-body systems, including tensor networks~\cite{vidal2003efficient,vidal2004efficient} and quantum Monte Carlo (QMC) methods~\cite{hastings2010measuring,humeniuk2012quantum,luitz2014improving,zhang2011entanglement,glasser2018neural}.
Recently, the representational power of neural network variational ans\"{a}tze has been successfully applied to study ground states and dynamics of many-body systems in both 1D and higher dimensions, and to reconstruct quantum states from experimental data~\cite{carleo2017solving, cai2018approximating,choo2019two,hartmann2019neural,yoshioka2019constructing,torlai2018neural, carrasquilla2019reconstructing,torlai2019integrating}. However, the numerical study of entanglement entropy for neural quantum states has received limited attention, and only calculations of $S_2$ for the two-layer restricted Boltzmann machine architecture have been demonstrated~\cite{deng2017quantum, torlai2018neural, glasser2018neural, torlai2019integrating}. To take advantage of state-of-the-art progress in machine learning and represent highly entangled  states more efficiently, deeper and more expressive network architectures have been introduced as ans\"{a}tze~\cite{cai2018approximating,sharir2020deep}. Exploiting such architectures for efficient entropy estimation has not yet been explored.

In this paper, we use quantum Monte Carlo methods to estimate generalized R\'{e}nyi entropies of quantum many body states parameterized by autoregressive neural networks.
A naive ``direct sampling'' approach performs well for small $n$ but fails for larger $n$, while an improved ``conditional sampling'' method  exploiting the autoregressive structure of the network ansatz outperforms direct sampling and enables calculations of higher-order R\'{e}nyi entropies in both 1D and 2D Heisenberg models for system sizes up to $N=256$ spins. Calculating R\'{e}nyi entropies $S_{n\geq 2}$ for integer $n$ allows for an approximation of the von Neumann entropy $S_1$ as well as extraction of the single copy entanglement $S_\infty$, which are difficult to access in traditional QMC.

\begin{figure}[t]
    \centering
    \includegraphics[width=1.0\columnwidth]{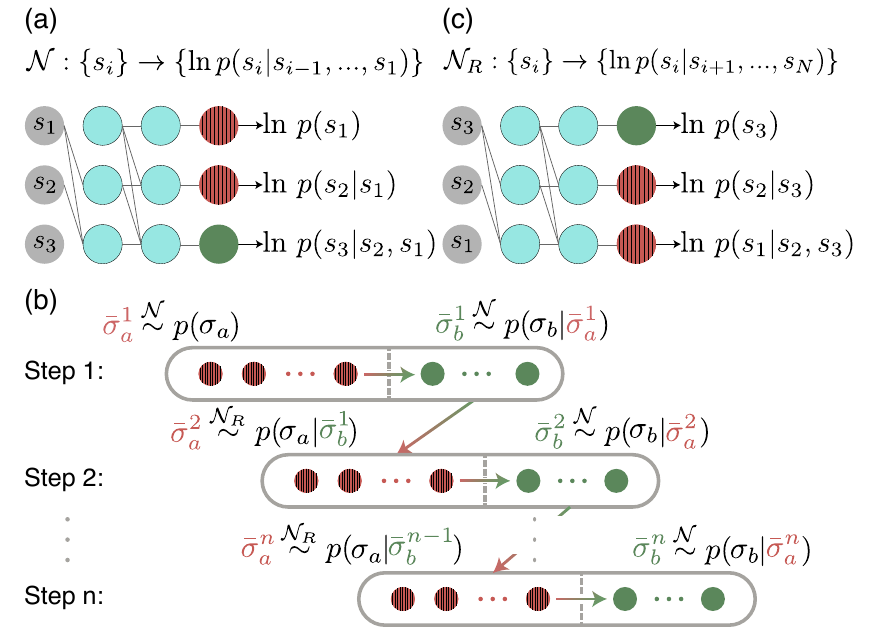}
    \caption{Network and sampling schematics. (a)~The autoregressive network $\mathcal{N}$  representing probability distribution $p(\sigma)$ consists of an input layer (gray) and hidden units (cyan outlined circles) with masked connections, followed by an output layer bipartitioned into subsystems $A$ (red, striped) and $B$ (green, solid). Network $\mathcal{N}$ takes an input spin configuration $\sigma=(s_1,s_2,s_3)$ and outputs the logarithm of $N$ conditional probabilities, which are summed to obtain $\ln[p({\sigma})]$. (b) Schematic illustration of conditional sampling sequence  $\bar{\sigma}_a^1 \rightarrow \bar{\sigma}_b^1 \rightarrow \bar{\sigma}_a^2 \rightarrow \bar{\sigma}_b^2 \rightarrow \cdots \rightarrow \bar{\sigma}_a^n \rightarrow \bar{\sigma}_b^n$. (c)~The reverse network $\mathcal{N}_R$ representing the same probability distribution as $\mathcal{N}$ is trained with a flipped ordering of the conditional probabilities.}
    \label{fig1}
\end{figure}

\section{Neural autoregressive quantum state}
A neural network may represent a quantum state of $N$ spins by taking a spin configuration as input and returning the corresponding amplitude and phase.
Concretely, the wavefunction in the computational basis $\sigma = (s_1, ..., s_N),s_i=\pm 1$ can be decomposed as
$\psi(\sigma) = \sqrt{p(\sigma)} e^{i \phi(\sigma)}$,
where $p(\sigma)$ and $\phi(\sigma)$ give the probability and phase for spin configuration $\sigma$. The network parameters are trained by minimization of the energy to represent a many-body ground state~\cite{carleo2017solving,sharir2020deep}. 
We choose an autoregressive network $\mathcal{N}$ to model $p(\sigma)$ [Fig.~\ref{fig1}(a)] \cite{sharir2020deep, wu2019solving}, and train a separate fully-connected network for the phase (see Appendix \ref{appendix:network}). Together, these comprise our neural autoregressive quantum state (NAQS). 

Autoregressive networks have several advantages for sampling applications compared to other neural quantum state architectures. They efficiently generate independent and identically distributed (iid) samples and directly output the normalized probability of each sample~\cite{germain2015made, oord2016pixel}, a substantial improvement over Markov chain Monte Carlo sampling required for e.g. restricted Boltzmann machines~\cite{carleo2017solving,glasser2018neural}. These features arise from the autoregressive structure: the output, a high-dimensional probability distribution, is expressed as a product of conditional probabilities $p(s_1, ..., s_N)~=~\prod_{i=1}^N p(s_i | s_{i-1},...,s_1)$.
Access to these conditionals allows direct generation of iid samples $\bar{\sigma}=(\bar{s}_1,...,\bar{s}_N)$ from the state distribution by sequentially drawing $\bar{s}_1\sim p(s_1),\bar{s}_2\sim p(s_2|\bar{s}_1),...,\bar{s}_N \sim p(s_N|\bar{s}_{N-1},...,\bar{s}_1)$.

\section{Calculating R\'{e}nyi entropies}
In this section, we develop two methods exploiting the autoregressive network structure to compute R\'enyi entanglement entropies of a neural quantum state. Samples from a network trained to represent target state $\psi$ may thereafter be used to compute observables such as  correlation functions~\cite{carleo2017solving, cai2018approximating,choo2019two,hartmann2019neural,yoshioka2019constructing,torlai2018neural, carrasquilla2019reconstructing,torlai2019integrating}. Estimating R\'{e}nyi entropies at integer orders $n \geq 2$ via the replica trick~\cite{calabrese2004entanglement, hastings2010measuring} is comparatively hard. The quantity $\tr{\rho_A^n}$ is computed explicitly as
\begin{align}
    &\tr{\rho_A^n} = \sum_{\{\sigma_a^k, \sigma_b^k\}} \langle \sigma_a^1, \sigma_b^1|\psi\rangle \langle \psi | \sigma_a^2, \sigma_b^1 \rangle...\langle \sigma_a^n, \sigma_b^n | \psi \rangle \langle \psi |  \sigma_a^1,\sigma_b^n \rangle
    \nonumber\\
    &= \sum_{\{\sigma^k_a, \sigma^k_b\}} \prod_{k=1}^{n} \psi(\sigma_a^k,\sigma_b^k) \psi^*(\sigma_a^{k+1},\sigma_b^k) \equiv \sum_{\bm{\sigma_a}, \bm{\sigma_b}} \Omega(\bm{\sigma_a}, \bm{\sigma_b}),
\end{align}
where $\sigma_a^{n+1} \equiv \sigma_a^1$ and the $n$ variables $\sigma_{a(b)}^k$ are computational basis vectors in $\mathcal{H}_{A(B)}$. For notational simplicity, we define $\Omega(\bm{\sigma_a}, \bm{\sigma_b})\equiv \prod_{k=1}^{n} \psi(\sigma_a^k,\sigma_b^k)  \psi^*(\sigma_a^{k+1},\sigma_b^k)$,
and let $\bm{\sigma_{a(b)}}\equiv\{\sigma^k_{a(b)},k=1,...,n\}$ be a set of $n$ basis vectors. 

\subsection{Direct sampling}
For the NAQS ansatz, a straightforward ``direct sampling'' (DS) estimator is
\begin{subequations}
\begin{alignat}{1}
    &\tr{\rho_A^n} = \ave{f_{\text{DS}}} = \ave{ \frac{\Omega(\bm{\sigma_a}, \bm{\sigma_b})}{P_{\text{DS}}(\bm{\sigma_a}, \bm{\sigma_b})} }_{(\bm{\sigma_a}, \bm{\sigma_b})\sim P_{\text{DS}}(\bm{\sigma_a}, \bm{\sigma_b})} \label{eq:tr_rho_n_ds} \\
    &P_{\text{DS}}(\bm{\sigma_a}, \bm{\sigma_b}) = \prod_{k=1}^n p(\sigma^k_a, \sigma^k_b). 
    \label{eq:ds_estimator}
\end{alignat}
\end{subequations}
In each Monte Carlo step, a batch of $n$ samples $\{(\bar{\sigma}^k_a, \bar{\sigma}^k_b),k=1,...,n\}$ is drawn independently from the state distribution $p(\sigma_a,\sigma_b)$, then permuted and recombined as $\{(\bar{\sigma}^{k+1}_a,\bar{\sigma}^k_b),k=1,...,n\}$ to evaluate the estimator $f_{\text{DS}}$; we average over $M$ batches.

\begin{figure}
    \centering
    \includegraphics[width=\columnwidth]{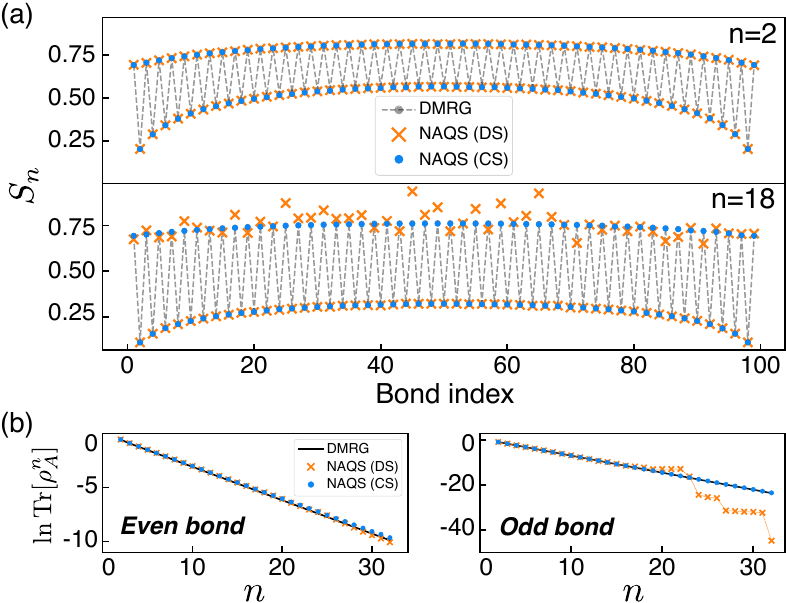}
    \caption{R\'{e}nyi entropies for 1D AFH ground state. (a) R\'{e}nyi entropies $S_2$ (top) and $S_{18}$ (bottom) for ground state of 1D AFH model with 100 spins and open boundary conditions. For $n=2$, DS (orange crosses) and CS (blue dots) data both match the DMRG results (gray dots; gray dashed line is a guide to the eye). For $n = 18$, direct sampling becomes comparatively noisier at the odd bonds. (b) $\ln \tr{\rho_A^n}$ for representative even bond index 50 and odd bond index 51 for $2 \leq n \leq 32$. Conditional sampling consistently gives close results to DMRG while direct sampling gets worse for larger $n$ at odd bonds.}
    \label{fig2}
\end{figure}

We benchmark direct sampling on a network trained to represent the ground state of a 1D antiferromagnetic Heisenberg (AFH) model $H = \sum_i \vec{s}_i \cdot \vec{s}_{i+1}$ for $N = 100$ spins.
As shown in Figure~\ref{fig2}(a), the direct sampling method (orange crosses) yields accurate results for $n=2$ compared with a DMRG computation (gray dots with dashed line)~\cite{tenpy}. However, for $n=18$ the entropy estimate across ``odd bonds'', which partition the system such that $A$ and $B$ have an odd number of spins, has larger variance. Figure~\ref{fig2}(b) shows that the agreement between direct sampling and DMRG is consistently close at even bonds but worsens by an order of magnitude at odd bonds for large $n$. This discrepancy can be explained in part by the spectrum of the reduced density matrix $\rho_A$: due to the SU(2) symmetry of the singlet ground state, the largest eigenvalue $\lambda_{\text{max}}$ of $\rho_A$ has degeneracy $g~=~2$ or $g=1$  at odd or even bonds respectively.

To understand why double degeneracy might cause increased variance across odd bonds for large $n$, we study a simple but illustrative example. Due to the $Z_2$ symmetry of the GHZ state $\ket{\psi_{\text{GHZ}}} = \frac{1}{\sqrt{2}} (\ket{\uparrow_A \uparrow_B} + \ket{\downarrow_A \downarrow_B})$, any bipartition results in a reduced density matrix $\rho_A$ with $g=2$. To calculate $\tr{\rho_A^n}$ using direct sampling, $M$ batches of $n$ independent samples are drawn, with each sample $(\bar{\sigma}^k_a, \bar{\sigma}^k_b) \in \{(\uparrow_A, \uparrow_B), (\downarrow_A, \downarrow_B)\}$. 
If $\bar{\sigma}^{k+1}_a$ and $\bar{\sigma}^k_b$ do not have the same spin configuration for all $n$ samples in the batch, the estimator $f_{\text{DS}} = 0$  since $\psi_{\text{GHZ}}(\uparrow_A,\downarrow_B)=\psi_{\text{GHZ}}(\downarrow_A,\uparrow_B) = 0$. 
The probability to independently draw $n$ aligned samples is $2^{1-n}$. 
Thus, the average over batches in Eq.~\ref{eq:tr_rho_n_ds} contains many terms equal to zero and the required number of samples grows exponentially with $n$.

\subsection{Conditional sampling}
To solve this variance problem and access higher-order R\'{e}nyi entropies, we propose an improved ``conditional sampling'' (CS) method, which generates correlated rather than independent samples within one batch.
Iteratively after drawing $\bar{\sigma}^k_a$, we sample $\bar{\sigma}^k_b \sim p(\sigma_b|\bar{\sigma}_a^k)$ followed by $\bar{\sigma}^{k+1}_a \sim p(\sigma_a | \bar{\sigma}_b^k)$, which generates the sample sequence $\bar{\sigma}_a^1 \rightarrow \bar{\sigma}_b^1 \rightarrow \bar{\sigma}_a^2 \rightarrow \bar{\sigma}_b^2 \rightarrow \cdots \rightarrow \bar{\sigma}_a^n \rightarrow \bar{\sigma}_b^n$ [Fig.~\ref{fig1}(b)]. Sampling in this order results in an estimator (see Appendix \ref{appendix:DS and CS}):
\begin{subequations}
\begin{alignat}{1}
    &\tr{\rho_A^n} = \ave{f_{\text{CS}}} = \ave{ \frac{\Omega(\bm{\sigma}_a, \bm{\sigma}_b)}{P_{\text{CS}}(\bm{\sigma}_a, \bm{\sigma}_b)} }_{(\bm{\sigma}_a, \bm{\sigma}_b)\sim P_{\text{CS}}(\bm{\sigma}_a, \bm{\sigma}_b)} \label{eq:tr_rho_n_cs} \\
    &P_{\text{CS}}(\bm{\sigma}_a, \bm{\sigma}_b) = p(\sigma^1_a)p(\sigma^1_b|\sigma^1_a)p(\sigma^2_a|\sigma^1_b)  \cdots p(\sigma^n_b|\sigma^n_a). \label{eq:cs_probability}
\end{alignat}
\end{subequations}
Assuming the predetermined sampling order of $\mathcal{N}$ is $\sigma_a \rightarrow \sigma_b$, then sampling in the other direction $\sigma_b \rightarrow \sigma_a$ requires a ``reverse network'' $\mathcal{N}_R$ which models the same probability distribution $p(\sigma)$ as $\mathcal{N}$, but outputs conditionals in the reverse order [Fig.~\ref{fig1}(c)]. We train $\mathcal{N}_R$ as a separate autoregressive network by minimizing its Kullback-Leibler divergence with $\mathcal{N}$ (see Appendix \ref{appendix:DS and CS}).

For the GHZ state, the sampling variance of the estimator $f_{\text{CS}}$ is zero because the correlated samples ensure alignment of $(\bm{\sigma_a}, \bm{\sigma_b})$ for all $n$. While the ground state of the 1D AFH model is comparatively more complex, this intuition generalizes:
Figure~\ref{fig2} shows that conditional sampling removes the larger variance at high $n$ compared to direct sampling, and the largest relative error of $S_{2\leq n \leq 32}$ compared with DMRG at all bonds is about 3.4\% due mainly to the network infidelity.
Heuristically, the success of conditional sampling here can be attributed to the existence of classical mutual information between regions $A$ and $B$ for bipartitions of the singlet ground state of the AFH model (see Appendix \ref{appendix:DS and CS}). 
\begin{figure}
    \centering
    \includegraphics[width=\columnwidth]{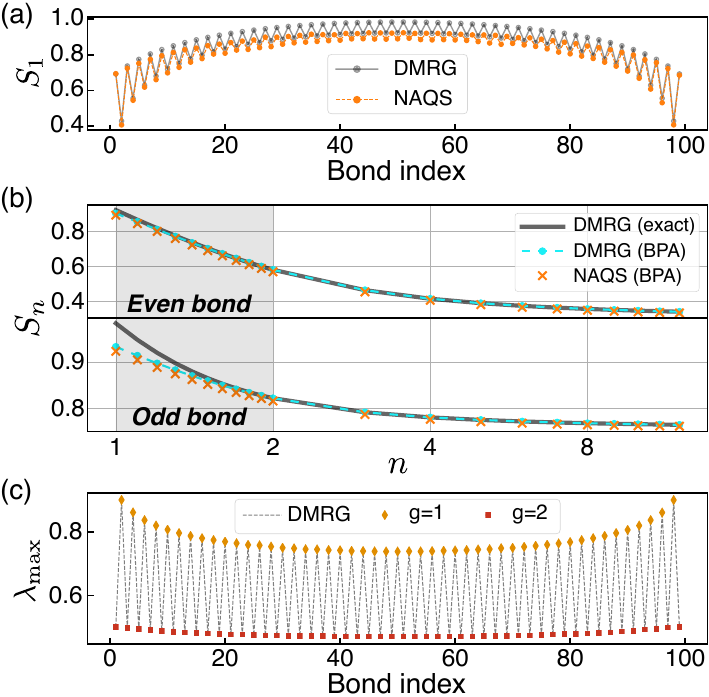}
    \caption{Extracted quantities. (a) The von Neumann entropy is approximated with
    BPA using R\'{e}nyi entropies up to $n_c=7$ obtained with direct sampling. 
    (b) R\'{e}nyi entropies $S_{1\leq n <2}$ are approximated with BPA for representative even bond index 50 and odd bond index 51. At odd bonds, the approximation error still exists even if exact $S_{2\leq n\leq 7}$ from DMRG are used in BPA, indicating that the systematic underestimate of $S_n$ is due to the cutoff error. (c) The largest eigenvalue $\lambda_\mathrm{max}$ and its degeneracy $g$ are extracted from a linear fit to the CS data for $\ln \tr{\rho_A^n}$, with $g$ restricted to an integer value.} 
    \label{fig3}
\end{figure}

\begin{figure}
    \centering
    \includegraphics[width=\columnwidth]{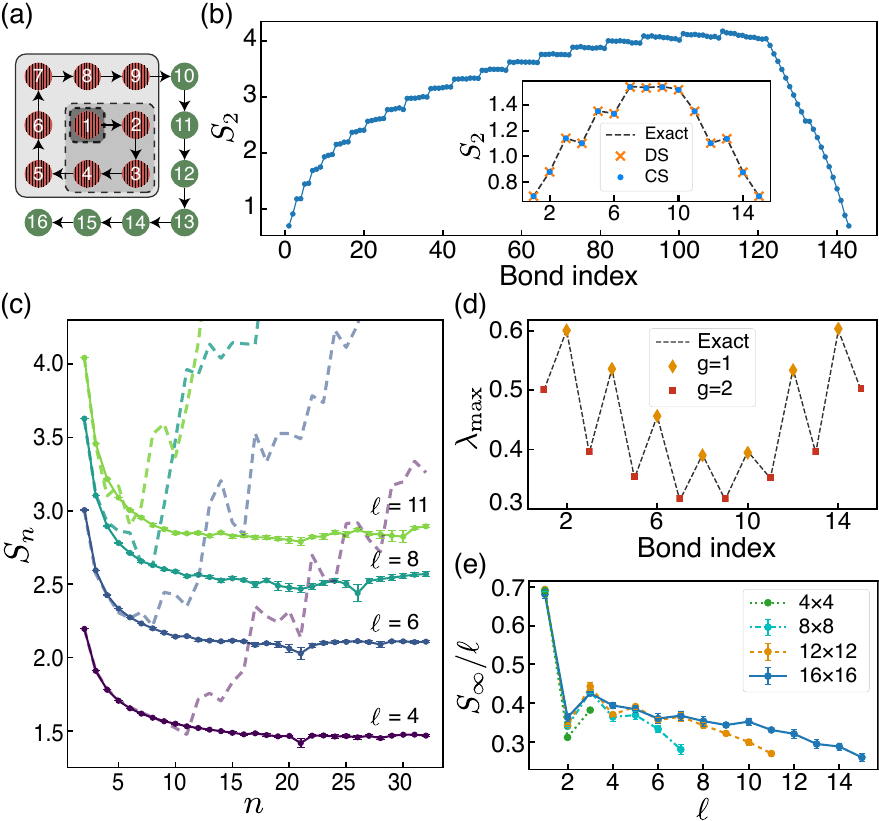}
    \caption{R\'{e}nyi entropies for 2D AFH ground states. (a)~Schematic of spiral ordering for network training and estimates of $S_n$. Lightest gray square indicates bipartition of the system into a $3\times 3$ subregion $A$ (red, striped) and remaining subregion $B$ (green, solid). Darker squares indicate partitions for estimating $S_n$ of $2\times2$ and $1\times1$ subregions.
    (b)~$S_2$ calculated with conditional sampling on a $12\times 12$ lattice with periodic boundary conditions. Inset: $S_2$ for $L=4$. DS (orange crosses) and CS (blue dots) data both match exact values (gray). (c)~$S_{2\leq n \leq 32}$ of $\ell\times \ell$ square regions for $L=12$ calculated with conditional sampling (solid) and direct sampling (dashed). (d) For $L=4$, $\lambda_{\text{max}}$ and $g$ extracted from CS data  match exact values. (e)~Single-copy entanglement $S_\infty(\ell)/\ell$ for $\ell\times \ell$ square regions in different 2D systems.
    }
    \label{fig4}
\end{figure}

\subsection{Approximating the von Neumann entropy}
R\'{e}nyi entropies at multiple integer orders $n\geq2$ contain strictly more information than a single order, which we harness to approximate the von Neumann entropy $S_1$. We compute the best polynomial approximation (BPA)
\begin{equation}
    S_1 = -\tr{\rho_A \log \rho_A} \approx \sum_{n=1}^{n_c} \alpha_n \tr{\rho_A^n},
\end{equation}
where $n_c$ is the cutoff polynomial degree.
The magnitude of the expansion coefficients grows exponentially with increasing $n$ (see Appendix \ref{appendix:vonNeumann}), so we choose $n_c=7$ to control the statistical error. Compared with the exact $S_1$ from DMRG, the polynomial estimator worsens near the center of the spin chain since small eigenvalues of $\rho_{A}$ contribute more to $S_1$ than $S_{n\geq2}$~[Fig.~\ref{fig3}(a)]. Non-integer R\'{e}nyi entropies $1<n<2$ can similarly be computed with a polynomial approximation, and like the von Neumann entropy are underestimated~ [Fig.~\ref{fig3}(b)]. This systematic error stems from the infidelity of the trained NAQS and from the cutoff at polynomial degree $n_c$; at $n_c=7$ the cutoff error dominates.

Conservatively, the required polynomial degree to maintain a controlled cutoff error for $S_1$ scales as $\sqrt{\chi}$, where $\chi$ is the rank of $\rho_A$ (see Appendix \ref{appendix:vonNeumann}). For the 1D AFH ground state, the bond dimension $\chi=100$ used in the DMRG calculation indicates that $n_c\approx 10$ should suffice. However, in 2D the required cutoff $n_c$ could be large since $\chi$ grows exponentially with the boundary size for area law states~\cite{vidal2003efficient}. Although it seems challenging to directly apply the BPA-based linear estimator to higher dimensions,
incorporating information from $S_{n\geq 2}$ provides a better estimate of $S_1$ than simply lower-bounding it by $S_2$.
Devising more robust estimators based on learning the nonlinear mapping between $S_{n \geq 2}$ and $S_1$ is a promising alternative to BPA~\cite{gray2018machine},
especially in higher dimensions where $S_1$ is generally inaccessible numerically except for states with special symmetries~\cite{mendes2019measuring}.

The entropies $S_{2\leq n\leq 7}$ used for BPA are generated by direct sampling, which in practice works surprisingly well for estimating low-order R\'{e}nyi entropies across all bonds. Direct sampling does not require a reverse network and is unrestricted by the ordering of the conditional probabilities in $\mathcal{N}$. As a result, the same samples can be reused to compute the entropy for different partitions whereas conditional sampling requires new samples for each bond. Moreover, direct sampling enables computations of the entanglement entropy for arbitrary bipartitions of the system and greater flexibility in the choice of network architecture. 
However, these advantages cannot overcome the exponential growth in the required number of samples with respect to $n$.

\subsection{Extracting the single copy entanglement}
In the limit as $n \rightarrow \infty$, it is necessary to estimate all $\ln \tr{\rho^n_A}$ with conditional sampling due to its reduced variance. In this regime, the main contribution to the entropy comes from the largest eigenvalue $\lambda_{\text{max}}$ of $\rho_A$, which yields the single copy entanglement $S_\infty  = -\ln \lambda_{\text{max}}$~\cite{eisert2005single}. We extract $\lambda_{\text{max}}$ from the slope and its degeneracy $g \in \mathbb{N}$ from the intercept of linear fits to $\ln \tr{\rho^n_A}$ with respect to $n$. We fit to contiguous subsets within the range $10 \leq n \leq 32$, and average over the results for all subsets with minimum length 10 (see Appendix \ref{appendix:lambdaMax}). 
The results are plotted in Figure~\ref{fig3}(c); the relative differences between the extracted $\lambda_{\text{max}}$ and DMRG results are within $0.8\%$. The marker colors indicate the fitted degeneracies, which match exact values.

\subsection{2D Heisenberg model}
We test both sampling methods in higher dimensions for the ground state of the 2D AFH model $H = \sum_{\langle i,j \rangle} \vec{s}_i \cdot \vec{s}_j$ on a $L \times L$ square lattice. The NAQS ansatz can efficiently represent quantum states in higher dimensions~\cite{wu2019solving,sharir2020deep}, which by contrast would require infeasibly large bond dimension for 1D tensor network states~\cite{vidal2003efficient}. To show area law scaling, we train our networks with a spiral ordering, enabling conditional sampling estimates of $S_n$ for regions of increasing area~[Fig.~\ref{fig4}(a)] (see Appendix \ref{appendix:network}).
Currently, the network training limits the largest accessible system size to $L=16$.
The R\'{e}nyi entropy $S_2$ for $L=12$ clearly shows the desired features of the spiral ordering choice [Fig.~\ref{fig4}(b)]. 
The boundary size of region $A$ increases in a stepwise pattern at each corner, which generates a series of entropy plateaus as expected for area law scaling. For $L=4$, we verify that $S_2$ calculated with both direct and conditional sampling agrees with exact diagonalization~[Fig.~\ref{fig4}(b) inset]~\cite{weinberg2017quspin}.

When estimating R\'{e}nyi entropies in 2D, conditional sampling still reduces the variance compared to direct sampling, especially at large $n$ where direct sampling fails to converge~[Fig.~\ref{fig4}(c)].
For $L=12$, we estimate $S_n$ reliably up to $n=10$ with largest relative error~2.6\% 
~[Fig.~\ref{fig4}(c)], and about $n=5$ for $L=16$ (see Appendix \ref{appendix:lambdaMax}). The increased error for larger system sizes is fundamentally related to the area law scaling of the Heisenberg ground state. The entanglement entropy scales linearly with the boundary size $L_A$ of $A$ 
and therefore $\tr{\rho_A^n} ~\sim \text{e}^{-n S_n} \sim \text{e}^{-nL_A}$, which indicates the trade-off between maximum order and system size.

The single-copy entanglement $S_\infty$ as the slope of $\ln\tr{\rho_A^n}$ still converges with conditional sampling despite the increased variance in individual $S_n$ (see Appendix \ref{appendix:lambdaMax}).
The extracted $\lambda_{\text{max}}$ and its degeneracy agree well with exact results for $L=4$~[Fig.~\ref{fig4}(d)]. For larger systems with $L=8, 12$, and 16, we extract $S_\infty(\ell)$ for $\ell \times \ell$ square regions $A$. In accordance with area law scaling, the normalized quantity $S_\infty(\ell)/\ell$ approaches a constant for $1\ll \ell \ll L$, especially for $L=16$ where finite size effects are suppressed~[Fig.~\ref{fig4}(e)]. Fitting the degeneracy, however, becomes more challenging at large system sizes due to the variance of the R\'{e}nyi entropy data.

\section{Discussion}
We have demonstrated two methods for calculating R\'{e}nyi entropies using neural autoregressive quantum states. The direct sampling method works relatively well for small $n$, and can readily be integrated with other network architectures~\cite{sharir2020deep,oord2018parallel} which could be particularly advantageous in higher dimensions or for incorporating symmetries.
Conditional sampling takes advantage of the autoregressive network to directly generate iid samples from the desired distribution~[Eq.~\ref{eq:cs_probability}] and significantly reduce the estimator variance, which reveals the potential of neural networks in designing more advantageous sampling schemes. Conditional sampling is currently the only existing method for computing $S_{n\geq2}$ efficiently for a neural quantum state.

Both sampling methods are independent of the system Hamiltonian and are straightforward to implement once the NAQS representation is trained, which can now be done using existing open-source software \cite{MADE_github,sharir2020deep}. 
Therefore, they can be directly applied to study R\'{e}nyi entropies of other quantum states including frustrated spin systems~\cite{choo2019two} and tomographic reconstructions of experimental data~\cite{torlai2019integrating}, and to investigate the entanglement structure of the NAQS architecture itself~\cite{deng2017quantum}. Moreover, conditional sampling performs competitively with more traditional QMC methods~\cite{hastings2010measuring,humeniuk2012quantum,luitz2014improving,zhang2011entanglement,glasser2018neural} and reaches similar or higher $n$ without introducing any problem-specific variance reduction tricks. 

In future work, improvements to the NAQS ansatz incorporating e.g. symmetries or using different generative models would be advantageous for training and testing our methods on larger system sizes or significantly increasing the sampling speed~\cite{oord2018parallel}. Proper combination of conditional sampling with ratio tricks~\cite{hastings2010measuring} could lead to better estimators with even smaller variance. More generally, designing sampling methods to suit a given network structure, or tailoring network architectures to complement a particular sampling scheme, may find broad use across a variety of Monte Carlo applications.

\begin{acknowledgments}
We thank Xiao-Liang Qi, Lei Wang, Yang Song, and Bohdan Kulchytskyy for useful discussions. This work was supported by the U.S. government through the Army Research Office (ARO) STIR Grant W911NF1910422. E.~J.~D. acknowledges support from the National Science Foundation Graduate Research Fellowship Program and from the Hertz Foundation. We thank Monika Schleier-Smith, Amir Safavi-Naeini, Ognjen Markovi\'{c} and Rishi Patel for feedback on the manuscript.

Z.~W. and E.~J.~D. contributed equally to this work.
\end{acknowledgments}

\appendix

\section{Comparing direct and conditional sampling}
\label{appendix:DS and CS}
In this section, we derive in greater detail the direct and conditional sampling estimators, including modifications to the CS estimator, and compare their variances. The R\'{e}nyi entropy $S_n$ for a bipartition of a pure state $\ket{\psi}\in \mathcal{H}_A \otimes \mathcal{H}_B$ requires calculation of $\tr{\rho_A^n}$, where $\rho_A$ is the reduced density matrix for subsystem $A$. For integer $n > 1$, $\tr{\rho_A^n}$ is accessible via the replica trick~\cite{hastings2010measuring}, which can be derived as
\begin{subequations}
\begin{alignat*}{1}
    \tr{\rho_A^n} & = \Tr_A \left[(\Tr_B \ket{\psi}\bra{\psi})^n \right] \\
    & = \Tr_A \sum_{\{\sigma_b^k\}} \inp{\sigma_b^1}{\psi} \inp{\psi}{\sigma_b^1} \inp{\sigma_b^2}{\psi} \cdots\\ &\hspace{18mm}\inp{\psi}{\sigma_b^2} \inp{\sigma_b^n}{\psi} \inp{\psi}{\sigma_b^n} \\
    & = \sum_{\{\sigma_a^k, \sigma_b^k\}} \inp{\sigma_a^1,\sigma_b^1}{\psi} \inp{\psi}{\sigma_a^2,\sigma_b^1} \inp{\sigma_a^2,\sigma_b^2}{\psi} \cdots \\ & \hspace{15mm}\inp{\psi}{\sigma_a^3,\sigma_b^2} 
    \inp{\sigma_a^n,\sigma_b^n}{\psi} \inp{\psi}{\sigma_a^1,\sigma_b^n} \\
    & = \sum_{\{\sigma_a^k, \sigma_b^k\}} \psi(\sigma_a^1,\sigma_b^1) \psi^*(\sigma_a^2,\sigma_b^1) \psi(\sigma_a^2,\sigma_b^2) \cdots \\ & \hspace{16mm}\psi^*(\sigma_a^3,\sigma_b^2)
    \psi(\sigma_a^n,\sigma_b^n) \psi^*(\sigma_a^1,\sigma_b^n)\\
    & \equiv  \sum_{\bm{\sigma_a}, \bm{\sigma_b}} \Omega(\bm{\sigma_a}, \bm{\sigma_b}),
\end{alignat*}
\end{subequations}
where the $n$ variables $\sigma_{a(b)}^k$ are computational basis vectors in $\mathcal{H}_{A(B)}$. For  notational simplicity, we have defined
\begin{equation}
    \Omega(\bm{\sigma_a}, \bm{\sigma_b})\equiv \prod_{k=1}^{n} \psi(\sigma_a^k,\sigma_b^k)  \psi^*(\sigma_a^{k+1},\sigma_b^k),
\end{equation}
with $\sigma_a^{n+1} \equiv \sigma_a^1$ and $\bm{\sigma_{a(b)}}\equiv\{\sigma^k_{a(b)},k=1,...,n\}$ a set of $n$ basis vectors. 

Two different Monte Carlo sampling schemes are proposed in the main text to estimate the sum in Eq.~S1. The first is direct sampling; here the trace $\Tr[\rho_A^n]$ is evaluated as an average over the estimator $f_\mathrm{DS}$, where
\begin{subequations}
\begin{alignat}{1}
    f_{\text{DS}} =& \frac{\Omega(\bm{\sigma_a}, \bm{\sigma_b})}{P_{\text{DS}}(\bm{\sigma_a}, \bm{\sigma_b})} = \prod_{k=1}^n \frac{\psi^*(\sigma_a^{k+1},\sigma_b^k)}{\psi^*(\sigma_a^k,\sigma_b^k)} \label{eq:estimator_ds},
\end{alignat}
\end{subequations}
with $P_{\text{DS}}(\bm{\sigma_a}, \bm{\sigma_b}) = \prod_{k=1}^n p(\sigma^k_a, \sigma^k_b)$, and where $p(\sigma_a,\sigma_b) = |\psi(\sigma_a,\sigma_b)|^2$ is the state distribution. The second method is conditional sampling; here the trace $\Tr[\rho_A^n]$ is evaluated as the expectation value of $f_\mathrm{CS}$, where
\begin{subequations}
\begin{alignat}{1}
    f_{\text{CS}} =& \frac{\Omega(\bm{\sigma_a}, \bm{\sigma_b})}{P_{\text{CS}}(\bm{\sigma_a}, \bm{\sigma_b})} = A(\bm{\sigma_a}, \bm{\sigma_b}) e^{i\Phi (\bm{\sigma_a}, \bm{\sigma_b})} \label{eq:estimator_cs} \\
    A(\bm{\sigma_a}, \bm{\sigma_b}) =& \sqrt{\frac{p(\sigma^1_a,\sigma^n_b) \prod_{k=1}^{n-1} p(\sigma^{k+1}_a)p(\sigma^k_b)}{P_{\text{CS}}(\bm{\sigma_a}, \bm{\sigma_b})}} \label{eq:A} \\
    \Phi (\bm{\sigma_a}, \bm{\sigma_b}) =& \sum_{k=1}^n \phi(\sigma^k_a,\sigma^k_b) - \sum_{k=1}^n \phi(\sigma^{k+1}_a,\sigma^k_b),
\end{alignat}
\end{subequations}
with
\begin{equation*}
    P_{\text{CS}}(\bm{\sigma_a}, \bm{\sigma_b}) = p(\sigma^1_a)p(\sigma^1_b|\sigma^1_a)p(\sigma^2_a|\sigma^1_b) p(\sigma^2_b|\sigma^2_a) \cdots p(\sigma^n_b|\sigma^n_a).
\end{equation*}
The estimator $f_{\text{DS}}$ can be calculated directly using outputs from the probability network $\mathcal{N}$, while $f_{\text{CS}}$ also requires outputs from the reverse network $\mathcal{N}_R$.

We now compare the variances of the estimators for the two sampling schemes. Since both are unbiased with the same mean value, comparing $\ave{f^2}$ will suffice. The trace $\tr{\rho_A^n}$ is real so we take the real part, yielding
\begin{subequations}
\begin{alignat}{1}
    \ave{f_{\text{DS}}^2} =& \sum_{\mathcal{D}(\bm{\sigma_a}, \bm{\sigma_b})} \left[ p(\sigma^1_a,\sigma^n_b) \prod_{k=1}^{n-1} p(\sigma^{k+1}_a,\sigma^k_b) \right] \cos^2 \Phi (\bm{\sigma_a}, \bm{\sigma_b})    \label{eq:ds2} \\
    \ave{f_{\text{CS}}^2} =& \sum_{\mathcal{D}(\bm{\sigma_a}, \bm{\sigma_b})} \left[ p(\sigma^1_a,\sigma^n_b) \prod_{k=1}^{n-1} p(\sigma^{k+1}_a)p(\sigma^k_b) \right] \cos^2 \Phi (\bm{\sigma_a}, \bm{\sigma_b}), 
    \label{eq:cs2}
\end{alignat}
\end{subequations}
where $\mathcal{D}(\bm{\sigma_a}, \bm{\sigma_b})$ is the summation range that satisfies $p(\sigma^k_a,\sigma^k_b)\neq 0,k=1,...,n$ as well as $p(\sigma^{k+1}_a,\sigma^k_b)\neq 0, k=1,...,n-1$.
The difference between Eqs.~\ref{eq:ds2} and \ref{eq:cs2} lies in the factors $p(\sigma_a)p(\sigma_b)$ and $p(\sigma_a,\sigma_b)$. If $\sigma_a$ and $\sigma_b$ are independent and hence $p(\sigma_a,\sigma_b) = p(\sigma_a)p(\sigma_b)$, then $\ave{f_{\text{DS}}^2}=\ave{f_{\text{CS}}^2}$. If there are correlations in the state probability distribution that result in classical mutual information across the partition, conditional sampling will automatically make use of this to avoid undesirable sample combinations with $\Omega (\bm{\sigma_a}, \bm{\sigma_b}) = 0$. 

This effect can be illustrated by comparing the variances for GHZ and product states
\begin{subequations}
\begin{alignat}{1}
    \ket{\psi_\mathrm{GHZ}} &= \frac{1}{\sqrt{2}} \left(\ket{\uparrow}^{\otimes N} + \ket{\downarrow}^{\otimes N} \right)\\
    \ket{\psi_\mathrm{P}} &= \left(\frac{1}{\sqrt{2}} (\ket{\uparrow} + \ket{\downarrow}) \right)^{\otimes N},
\end{alignat}
\end{subequations}
for which the estimators are simple enough to evaluate analytically. As stated in the main text, estimating $S_n$ for the highly-correlated GHZ state strongly benefits from generating correlated samples. In fact, the variance $\text{Var}[f_{\text{CS}}] =  4^{1-n} - 4^{1-n} = 0$
for all orders $n$. By contrast, for direct sampling $\text{Var}[f_{\text{DS}}] = 2^{1-n}- 4^{1-n}$
and the required number of samples to reach fixed variance scales exponentially in $n$ as $\text{Var}[f_{\text{DS}}]/\ave{f_{\text{DS}}}^2 \approx 2^{n-1}$. However, conditional sampling yields no benefit for the uncorrelated product state $\ket{\psi_\mathrm{P}}$,
since $\text{Var}[f_{\text{DS}}] = \text{Var}[f_{\text{DC}}] = 1$.

\begin{figure}
    \centering
    \includegraphics[width=0.48\textwidth]{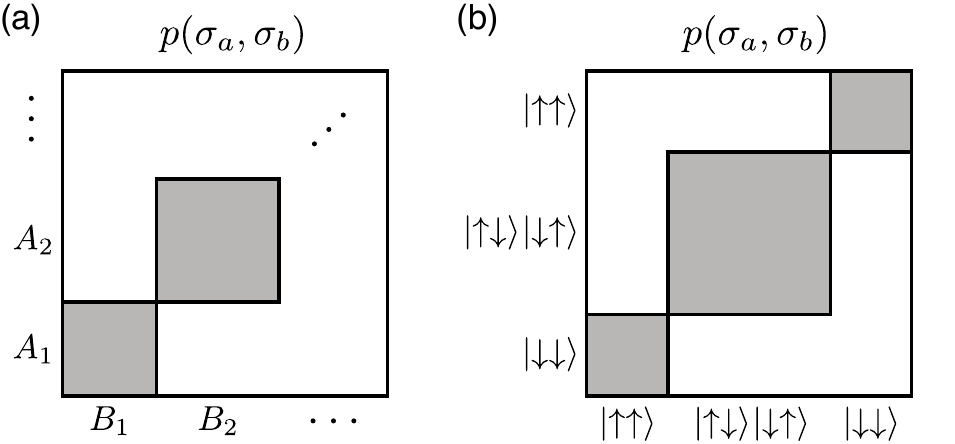}
    \caption{(a) Schematic for a block diagonal structure of the state distribution in the computational basis for a fixed bipartition. White regions have probability zero. (b) Example decomposition into blocks for a cut through the center of the ground state of 4-spin AFH model. }
    \label{fig:block}
\end{figure}

We generalize the argument beyond these two specific cases to states whose probability distribution is ``block diagonal", leaving the completely general case for future work. We call the distribution $p(\sigma_a,\sigma_b)$ block diagonal if $\Omega_A$ and $\Omega_B$, the sampling spaces for $\sigma_a$ and $\sigma_b$, can be decomposed into the union of $C$ non-overlapping subsets~[Fig.~\ref{fig:block}(a)]
\begin{equation}
    \Omega_A = \bigcup_{i=1}^C A_i,\quad \Omega_B = \bigcup_{i=1}^C B_i
\end{equation}
such that
\begin{equation}
    \forall i \neq m, \quad p(\sigma_a \in A_i, \sigma_b \in B_m) = 0.
\end{equation}

The distribution for the ground state of the antiferromagnetic Heisenberg (AFH) model on a  bipartite lattice is block diagonal. The ground state is a spin singlet with total spin $S = 0$ and total $S_z = 0$~\cite{marshall1955antiferromagnetism,lieb1962ordering}. 
Thus, $p(\sigma_a,\sigma_b)$ can be decomposed into the following blocks:
\begin{equation}
    A_i = \{ \ket{\sigma_a^z=j} \},B_i = \{ \ket{\sigma_b^z=-j} \},
\end{equation}
where
\begin{equation*}
    \begin{cases}
        j=0,\pm 1,..., \pm \frac{1}{2} \min \{n_A,n_B\} & \text{if $n_A, n_B$ are even}\\
        j=\pm \frac{1}{2},\pm \frac{3}{2},..., \pm \frac{1}{2} \min \{n_A,n_B\} & \text{if $n_A, n_B $ are odd}, \\
    \end{cases}
\end{equation*}
with $n_A$ and $n_B$  the number of spins in regions $A$ and $B$.
As a concrete example, the structure of the ground state distribution for a cut through the center of a 4-spin AFH ground state is shown in Fig.~\ref{fig:block}(b). Here, the sampling space is decomposed into 3 blocks as
\begin{alignat}{3}
    A_1 &= \{\ket{\uparrow \uparrow}\}, A_2 &= \{\ket{\uparrow \downarrow},\ket{\downarrow \uparrow}\},A_3 &= \{\ket{\downarrow \downarrow}\} \\
    B_1 &= \{\ket{\downarrow \downarrow}\}, B_2 &= \{\ket{\uparrow \downarrow},\ket{\downarrow \uparrow}\},B_3 &= \{\ket{\uparrow \uparrow}\} .
\end{alignat}

Each of the $n$ independent samples $(\sigma^k_a,\sigma^k_b), k=1,...,n$ generated by direct sampling in one batch falls into any one of the $C$ blocks, which is undesirable because samples from different blocks result in $f_{\text{DS}}=0$. Conversely, $f_{\text{DS}}(\bm{\sigma_a}, \bm{\sigma_b})\neq 0$ if all samples in one batch are contained within the same block, which happens with probability  $\sim 1/C^{n-1}$ if we assume each block has probability $1/C$. Conditional sampling, by contrast, guarantees that all samples stay in the same block with unity probability due to the correlation between samples.

Although direct sampling fails to converge at large $n$ due to this exponential growth of variance, it still performs relatively well at small $n$. Specifically, Fig.~\ref{fig4}(c) reveals that low-order entropies are faithfully calculated with direct sampling for a $12\times 12$ system. Moreover, direct sampling reuses the same samples for the entropy calculation across different bonds, resulting in a speed-up proportional to the total number of partitions. For example, computing $S_2$ up to fixed accuracy with direct sampling across all bonds for $L=12$ [Fig.~\ref{fig4}(b)] is about twice as fast as conditional sampling.

We add here a few comments related to the seminal QMC work by Hastings et al.~\cite{hastings2010measuring}. First, both estimators $f_{\text{DS}}$ and $f_{\text{CS}}$ are symmetric for regions $A$ and $B$, ensuring that estimates of $\tr{\rho_A^n}$ and $\tr{\rho_B^n}$ have the same mean and variance. Therefore, the 1D results presented in the main text avoid the problem of monotonically increasing variance with respect to the size of $A$, as seen with a naive swap method in Fig.~2 of Ref.~\cite{hastings2010measuring}. Second, their ratio trick cannot be directly applied in our work since the NAQS uses an orthogonal $S_z$ basis and the wavefunction in general is complex. This differs significantly from their use of the valence bond basis where the wavefunction is positive for the Heisenberg model on a bipartite lattice. Combining the ratio trick with neural quantum states and the methods presented here is an interesting direction for future work.

\subsection{Batch sizes for QMC results in the main text}
Since the variance of the CS estimator is small, we are able to achieve reasonable results with fewer samples than required for the DS estimator to converge. For some cases at high $n$, e.g. Fig.~2(b) and Fig.~4(c) in the main text, we do not even try to get enough DS data to converge to a reasonable value as it would be unrealistic. For Fig.~2(a-b) the number of batches is about $M=5\times 10^5$ for conditional sampling and $M=1\times 10^6$ for direct sampling. For Fig.~3(a-b) we have $M=4\times 10^8$ with direct sampling and (c) $M=5\times 10^5$ with conditional sampling. We choose direct sampling to calculate $S_n$ for BPA since each batch of samples can be reused to compute the entropies for all the bipartitions as in Fig.~3a, while conditional sampling would require new samples for each bipartition. Hence, obtaining a fixed number of samples for each bond is faster with direct sampling by a factor equal to the number of bonds across which the entanglement entropy is estimated. For Fig.~4(b), $S_2$ for the $12\times 12$ system is estimated using conditional sampling with $M=3\times 10^5$. For the inset, $M=5\times 10^6$ for CS and $M=3\times 10^7$ for DS. For Fig.~4(c), the batch size for conditional sampling ranges from $8\times 10^6$ to $1\times 10^8$ and for direct sampling $M=5\times 10^6$. Fig.~4(d) uses CS data with $M=5\times 10^6$. For Fig.~4(e) we use conditional sampling and the number of batches ranges from $2\times 10^6$ to $1\times 10^8$ depending on the variance of each bipartition.

\subsection{Modifications to CS Estimator}
Here, we show how to modify the CS estimator to account for imperfect fidelity between the networks $\mathcal{N}$ and $\mathcal{N}_R$, and to further control its variance. After training the first network $\mathcal{N}$ to represent probability $p(\sigma_a, \sigma_b)$, we train the reverse network $\mathcal{N}_R$ by minimizing the Kullback-Leibler (KL) divergence 
\begin{equation}
    D_{\text{KL}}(\mathcal{N}_R) = \sum_{\sigma_a,\sigma_b} p(\sigma_a,\sigma_b) \ln \frac{p(\sigma_a,\sigma_b)}{p_R(\sigma_a,\sigma_b)},
\end{equation}
where $p_R(\sigma_a,\sigma_b)$ is the probability distribution represented by the  reverse network. Ideally the divergence would be zero, meaning $p(\sigma_a,\sigma_b) = p_R(\sigma_a,\sigma_b)$; however, in our experiments there always exists a small infidelity on the order of~$\sim 0.5\%$. Taking this infidelity into account, the sampling weight for conditional sampling becomes
\begin{equation}
    \begin{split}
        P_{\text{CS}}(\bm{\sigma_a}, \bm{\sigma_b}) =& p(\sigma^1_a)p(\sigma^1_b|\sigma^1_a)p_R(\sigma^2_a|\sigma^1_b) p(\sigma^2_b|\sigma^2_a) \cdots \\
        & p_R(\sigma^n_a|\sigma^{n-1}_b) p(\sigma^n_b|\sigma^n_a),
    \end{split}
\end{equation}
and the estimator also has to be modified correspondingly as (compare to Eq.~\ref{eq:A})
\begin{equation}
    \label{eq:A2}
    \begin{split}
        A(\bm{\sigma_a}, \bm{\sigma_b}) =& \frac{ \sqrt{\prod_{i=1}^{n} p(\sigma^{i}_a,\sigma^i_b) \prod_{i=1}^{n} p(\sigma^{i+1}_a,\sigma^i_b) }}{P_{\text{CS}}(\bm{\sigma_a}, \bm{\sigma_b})} \\
        =& \sqrt{\frac{p(\sigma_a^1,\sigma_b^n)}{p(\sigma_a^1,\sigma_b^1)}} \prod_{i=1}^{n-1} \frac{\sqrt{p(\sigma^{i+1}_a,\sigma^i_b)}}{p_R (\sigma_a^{i+1}|\sigma^i_b)} \prod_{i=2}^n \sqrt{\frac{p(\sigma_a^i)}{p(\sigma_b^i|\sigma^i_a)}}.
    \end{split}
\end{equation}
This estimator is unbiased, and therefore the imperfection of the reverse network has no effect on the value of $\tr{\rho_A^n}$. However, mismatch between $p$ and $p_R$ increases the variance and therefore we still want to maximize the fidelity.

Another contribution to the variance comes from the factor of $p(\sigma_a^1,\sigma_b^n)$ in Eq.~\ref{eq:A2}. Because this factor is not sampled, it
can change $A(\bm{\sigma_a}, \bm{\sigma_b})$ by orders of magnitude for different sample sequences. To control this variance, we introduce a simple trick to include $p(\sigma_a^1,\sigma_b^n)$ in the sampling weight. After generating the sequence $\bar{\sigma}_a^1 \rightarrow \bar{\sigma}_b^1 \rightarrow \bar{\sigma}_a^2 \rightarrow \bar{\sigma}_b^2 \rightarrow \cdots \rightarrow \bar{\sigma}_b^{n-1} \rightarrow \bar{\sigma}_a^n$ as before, we sample $\bar{\sigma}_b^n \sim p(\sigma_b | \bar{\sigma}_a^n)$ or $\bar{\sigma}_b^n \sim p(\sigma_b | \bar{\sigma}_a^1)$ with probability 0.5. The sampling weight becomes
\begin{equation}
    \begin{split}
        P_{\text{CS}}(\bm{\sigma_a}, \bm{\sigma_b}) = & p(\sigma^1_a)p(\sigma^1_b|\sigma^1_a)p_R(\sigma^2_a|\sigma^1_b) p(\sigma^2_b|\sigma^2_a) \cdots \\
        &   p_R(\sigma^n_a|\sigma^{n-1}_b) \frac{p(\sigma^n_b|\sigma^n_a) + p(\sigma^n_b|\sigma^1_a)}{2},
    \end{split}
\end{equation}
which effectively contains $p(\sigma_a^1,\sigma_b^n)$. The estimator is correspondingly modified as
\begin{equation}
    \begin{split}
        A(\bm{\sigma_a}, \bm{\sigma_b}) =& \sqrt{\frac{p(\sigma_a^1,\sigma_b^n)}{p(\sigma_a^1,\sigma_b^1)}} \prod_{i=1}^{n-1} \frac{\sqrt{p(\sigma^{i+1}_a,\sigma^i_b)}}{p_R (\sigma_a^{i+1}|\sigma^i_b)} \\
        & \prod_{i=2}^n \sqrt{\frac{p(\sigma_a^i)}{p(\sigma_b^i|\sigma^i_a)}} \frac{p(\sigma^n_b | \sigma^n_a)}{\frac{1}{2} [p(\sigma^n_b | \sigma^n_a) + p(\sigma^n_b | \sigma^1_a)]}.
    \end{split}
\end{equation}

\section{Implementing symmetries in entropy calculation}
\label{appendix:symmetries}
Many-body ground states often exhibit symmetries like rotational or translational invariance, and identifying and enforcing these problem-dependent symmetries while training neural quantum states can greatly improve the accuracy of the resulting representation. A symmetry can be enforced by applying all symmetry transformations to the input spin configuration, then taking the average of the network outputs. For example, to add $Z_2$ symmetry to the network $\mathcal{N}$, we use
\begin{equation}
    P_{Z_2}(\sigma) = \frac{1}{2} \left( P_{\mathcal{N}}(\sigma) + P_{\mathcal{N}}(-\sigma) \right)
\end{equation}
as the state distribution instead of the direct output from the network $P_{\mathcal{N}}(\sigma)$. To generate samples from $P_{Z_2}(\sigma)$, we sample $\bar{\sigma}$ from $P_{\mathcal{N}}(\sigma)$ and then randomly flip $\bar{\sigma}$ to $-\bar{\sigma}$ with probability 0.5. Similarly, for translational invariance
\begin{equation}
    \begin{split}
        & P_{\text{trans}} (s_1, .., s_N) = \\
        & \frac{P_{\mathcal{N}}(s_1, .., s_N) + P_{\mathcal{N}}(s_2, .., s_N,s_1) + .. + P_{\mathcal{N}}(s_N, s_1, .., s_{N-1})}{N}
    \end{split}
\end{equation}
can be used as the state distribution. To generate samples from $P_{\text{trans}}$, samples are first generated from $P_{\mathcal{N}}(\sigma)$ and then one of the $N$ cyclic permutations is applied with probability $1/N$.

After obtaining the ground state with symmetries implemented as above, the entropy calculation using conditional sampling has to be modified accordingly. Challenges include calculating the conditional probabilities for the symmetrized state distribution and generating samples from these conditional probabilities. For $Z_2$ symmetry, 
both $P_{Z_2}(\sigma_a)$ and $P_{Z_2} (\sigma_b |\sigma_a)$ required for the conditional sampling estimator $f_{\text{CS}}$ can be derived as
\begin{equation}
    \begin{split}
        P_{Z_2}(\sigma_a) &= \frac{1}{2}(P_{\mathcal{N}}(\sigma_a) + P_{\mathcal{N}}(-\sigma_a)) \\
        P_{Z_2} (\sigma_b |\sigma_a) &= \frac{P_{\mathcal{N}}(\sigma_a) P_{\mathcal{N}} (\sigma_b |\sigma_a) + P_{\mathcal{N}}(-\sigma_a) P_{\mathcal{N}} (-\sigma_b |-\sigma_a)}{P_{\mathcal{N}}(\sigma_a) + P_{\mathcal{N}}(-\sigma_a)} .
    \end{split}
\end{equation}
To generate samples $\bar{\sigma}_a \sim P_{Z_2}(\sigma_a)$, we sample directly from $P_{\mathcal{N}}(\sigma_a)$ and then flip $\bar{\sigma}_a$ to $-\bar{\sigma}_a$ with probability 0.5. For $\bar{\sigma}_b \sim P_{Z_2} (\sigma_b |\bar{\sigma}_a)$, with probability $p=P_{\mathcal{N}}(\bar{\sigma}_a)/(P_{\mathcal{N}}(\bar{\sigma}_a) + P_{\mathcal{N}}(-\bar{\sigma}_a))$ we sample directly as $\bar{\sigma}_b \sim P_{\mathcal{N}} (\sigma_b |\bar{\sigma}_a)$ and with probability $1-p$ we sample first from $\bar{\sigma}_b \sim P_{\mathcal{N}} (\sigma_b |-\bar{\sigma}_a)$ and then flip the sample $\bar{\sigma}_b \rightarrow -\bar{\sigma}_b $.

Unfortunately, for translational invariance a similar trick cannot be applied since even for $P_{\text{trans}} (s_1)$ the calculation involves intractable terms like $\sum_{s_2,...,s_N} P_{\mathcal{N}}(s_2, ..., s_N,s_1)$. We therefore only enforce $Z_2$ symmetry when training the networks. By contrast, enforcing translational symmetry for direct sampling is not a problem since neither sample generation nor estimator calculation require $P_{\text{trans}} (\sigma_b | \sigma_a)$.

\section{Best polynomial approximation of von Neumann entropy}
\label{appendix:vonNeumann}
The von Neumann entropy $S_1 = -\tr{\rho \ln \rho}$ is calculated as a sum $-\sum_i \lambda_i \ln \lambda_i$ where $\lambda_i \in (0,1]$ are the eigenvalues of $\rho$. If we find a polynomial approximation for $f(x) = -x\ln x$ over the range $(0,1]$ such that $f(x) \approx \sum_{n} \alpha_n x^n$, then the Von Neumann entropy can be approximated as
\begin{equation}
    \begin{split}
        & -\tr{\rho \ln \rho} = -\sum_i \lambda_i \ln \lambda_i  \\
        \approx & \sum_i \sum_{n} \alpha_n \lambda_i^n = \sum_{n} \alpha_n \sum_i \lambda_i^n = \sum_{n} \alpha_n \tr{\rho^n},
    \end{split}
\end{equation}
where $\tr{\rho^n}$ are directly measurable from NAQS sampling.

\begin{figure}
    \centering
    \includegraphics[width=0.48\textwidth]{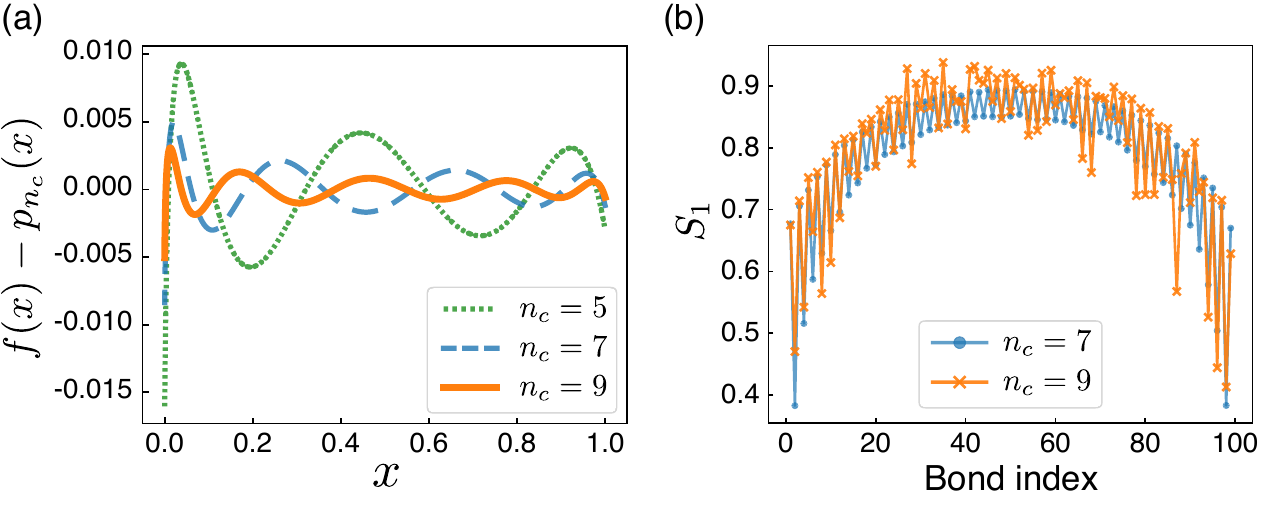}
    \caption{(a) Error for best polynomial approximation of $f(x) = -x\ln x$ with different cutoff degrees $n_c$. (b) Polynomial approximation for $S_1$ at cutoff degree $n_c=9$ computed with direct sampling has larger variance compared with $n_c=7$, which is also plotted in Fig.~3(a) of the main text.}
    \label{fig:polyfit}
\end{figure}

The best polynomial approximation (BPA) for $f(x)$ is defined as
\begin{equation}
    p_n^* = \min_{p_n \in \mathcal{P}_n} ||f-p_n|| = \min_{p_n \in \mathcal{P}_n} \max_{x \in (0,1]} |f(x)-p_n(x)|,
\end{equation}
where $\mathcal{P}_n$ is the set of all polynomials with degree $n$. For most functions $f(x)$, deriving an explicit formula for $p_n^*$ is impossible and also unnecessary~\cite{trefethen2013approximation,devore1993constructive}. Instead, we find a near-best solution where the approximation error is provably well-controlled compared to $p_n^*$~\cite{trefethen2013approximation}. One way to obtain these near-best polynomials is to expand $f(x)$ in terms of Chebyshev polynomials
\begin{equation}
    \begin{split}
        T_{0}(x)&=1\\
        T_{1}(x)&=x\\
        T_{n+1}(x)&=2xT_{n}(x)-T_{n-1}(x),
    \end{split}
\end{equation}
which have broad applications in approximation theory~\cite{trefethen2013approximation,devore1993constructive}. We apply an affine transformation and define the Chebyshev polynomials on [0,1] as $\hat{T}_n(x) = T_n(2x-1)$. We then expand $f(x)$ using $\hat{T}_n(x)$ as a function basis:
\begin{equation}
    f(x)=-x\ln x \approx p_{n_c}(x) = \sum_{k=0}^{n_c} a_k \hat{T}_k(x),
\end{equation}
where $n_c$ is the cutoff degree and the expansion coefficients are given by~\cite{wihler2014computing}
\begin{align}
    & a_0 = \ln 2 - \frac{1}{2}, \quad a_1 = \ln 2 - \frac{3}{4} \\
    & a_k = \frac{(-1)^{k+1}}{k(k^2-1)}, \quad k \geq 2 ,
\end{align}
and the error of the expansion is bounded by
\begin{equation}
    \max_{x \in (0,1]} |f(x)-p_{n_c}(x)| \leq \frac{1}{2n_c(n_c+1)}.
\end{equation}
We plot the polynomial approximation error for different cutoff orders in Fig.~\ref{fig:polyfit}(a), which clearly shows a reduction in error for larger $n_c$.

If the rank of $\rho$ is $\chi$, then the total estimation error for the von Neumann entropy is bounded by
\begin{equation}\label{eq:vn_bound}
    \left| \sum_{i=1}^{\chi} f(\lambda_i) - p_{n_c}(\lambda_i) \right| \leq \frac{\chi}{2n_c(n_c+1)} .
\end{equation}
To guarantee convergence, the polynomial degree $n_c$ has to scale as $\sqrt{\chi}$. For area-law states in 1D, $\chi$ does not scale with the size of the subsystem. The rank $\chi \sim 100$ used in the DMRG calculation suggests $n_c \approx 10$ should suffice, which is close to the maximum degree 7 that we used. In higher dimensions, since $\chi$ grows exponentially with the boundary size of the subsystem even for area-law states, the required polynomial degree could be very large. On the other hand, the bound in Eq.~\ref{eq:vn_bound} is completely general and may not be saturated for all physical states.

\begin{figure*}
    \centering
    \includegraphics[width=0.95\textwidth]{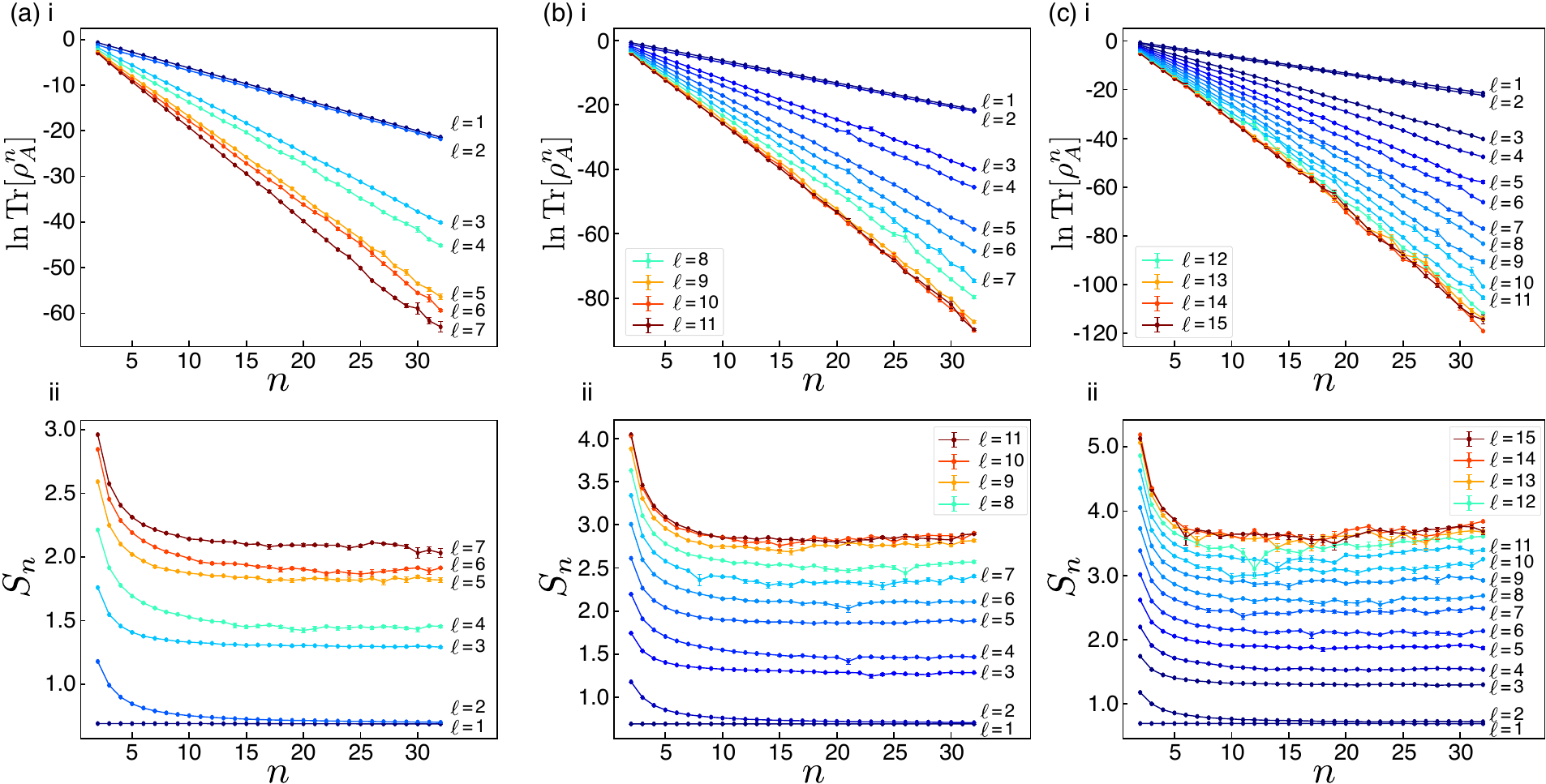}
    \caption{(i) $\ln \tr{\rho_A^n}$ and (ii) corresponding R\'{e}nyi entropy $S_n$ for $\ell \times \ell$ regions $A$  in (a) $8\times 8$ (b) $12\times 12$ and (c) $16\times 16$ systems. All results here are calculated with conditional sampling. $S_\infty$ in Fig.~4(e) of the main text is extracted from fits to the data in (i).}
    \label{fig:2D_results}
\end{figure*}

To see what happens to the polynomial approximation at higher cutoffs $n_c$, we calculate the absolute value of the leading coefficient of $p_{n_c}(x)$. From the recurrence relation of Chebyshev polynomials as well as the affine transformation, we have
\begin{equation}
    |\alpha_{n_c}| = \frac{2^{2{n_c}-1}}{n_c(n_c^2-1)},
\end{equation}
which is dominated by the exponential growth at large $n_c$. Therefore, any small statistical error in $\tr{\rho^n}$ leads to very large error in the estimated Von Neumann entropy, which makes the polynomial approximation sensitive to statistical noise. Fig.~\ref{fig:polyfit}(b) shows the $S_1$ calculation for the ground state of the 1D AFH model for cutoff degrees $n_c = 7$ and $9$. At $n_c = 7$, the variance is controlled and this data is plotted in Fig.~3(a) of the main text. However, at $n_c = 9$ the variance is significantly higher and leads to a worse approximation. The two competing requirements for large cutoff and small statistical error need to be balanced.

\section{Extracting $\lambda_{\text{max}}$ from R\'{e}nyi entropies}
\label{appendix:lambdaMax}
Assuming $\lambda_{\text{max}}$, the maximum eigenvalue of $\rho_A$, has a finite spectral gap 
from all other eigenvalues of $\rho_A$ and its degeneracy is $g$, then for large $n$
\begin{equation}
    \ln \tr{\rho_A^n} = \ln \sum_{i} \lambda_i^n \approx \ln (g \lambda_{\text{max}}^n ) = \ln g + n \ln \lambda_{\text{max}}.
\end{equation}
We can extract $\lambda_{\text{max}}$ from the slope and $g$ from the intercept of a linear fit to $\ln \tr{\rho_A^n}$ in the large $n$ range, with $g$ restricted to integer values. 
This works well in 1D and the $4\times 4$ system in 2D, and both yield results matching exact calculations [Fig.~3(c), Fig.~4(d) in main text]. 

Here, we plot and discuss the CS data used to extract the single-copy entanglement $S_\infty = -\ln \lambda_\mathrm{max}$ plotted in Fig.~4(e) in the main text. At these larger system sizes in 2D, we observe an increased variance in $S_n$ attributable to the exponentially small values $\Tr[\rho_A^n] \sim e^{-n L_A}$. The full datasets for $8\times8, 12\times12$, and $16\times16$ systems are shown in Fig.~\ref{fig:2D_results}. Despite the increased variance in $S_n$, the data for $\ln \Tr[\rho_A^n]$ still behave linearly and it is reasonable to fit a slope to them. The variance of the data also makes the slope $\ln \lambda_{\text{max}}$ slightly depend on the range of $n$ that we choose, and we therefore average over the results for all contiguous subsets with minimum length 10 within the range $10\leq n \leq 32$.

\section{Network structures and training}
\label{appendix:network}
The NAQS contains an autoregressive probability network $\mathcal{N}$  and a separate phase network which is fully-connected. The probability network is based on MADE~\cite{germain2015made,wu2019solving}, which uses masked connections to preserve the autoregressive property. The fully-connected phase network has the same depth (total number of layers) and width (number of channels of each hidden layer) as the probability network, but only has a single output instead of $N$ outputs. For training the ground states of the Heisenberg model, we usually choose a depth of 3 to 4,
and the width ranges from 4 to 16 depending on the system size as well as the network depth. We choose very similar structures for the reverse networks $\mathcal{N}_R$. 

The minimization of both energy and KL divergence are done with the Adam optimizer~\cite{kingma2014adam}. We usually start with learning rate $10^{-3}$ and batch size 1000 for about 5000 steps, then gradually increase the batch size to 10000 until the optimization stops improving. Finally, we iterate between reducing the learning rate and increasing the batch size by a factor of 3 to 10.

Specific properties of the Heisenberg ground state can be leveraged to achieve faster training and benefit the entropy calculation. To accelerate ground state training of the $12\times 12$ and $16\times 16$ system, we only train the probability network and circumvent the phase network by directly applying the Marshall sign rule~\cite{marshall1955antiferromagnetism}. For all other system sizes, both phase network and probability network are trained. As shown in Sec.~\ref{appendix:DS and CS}, the $S_z=0$ property of Heisenberg ground state leads to a block diagonal $p(\sigma_a,\sigma_b)$ which reduces the variance of conditional sampling. However, in our training of the $8\times 8$, $12\times 12$ and $16\times 16$ systems, the relative errors for the ground state energy are only on the order of $10^{-3}$ and therefore the networks can still generate samples with $S_z \neq 0$. Those samples lead to hopping between different diagonal blocks of $p(\sigma_a,\sigma_b)$ and cause larger variance at higher $n$. To solve this problem, we implement an extra penalty term for all $S_z\neq 0$ samples from the networks, which reduces the probability of $S_z\neq 0$ samples to around $0.1\%$ after training. Empirically, this extra penalty term also helps in avoiding local energy minima and leads to networks with lower energy. For the 1D case as well as $4\times 4$ in 2D, the extra penalty is not necessary since the relative energy errors are around $10^{-5} \sim 10^{-4}$ and $S_z=0$ is automatically satisfied with high probability.

\begin{figure}
    \centering
    \includegraphics[width=0.48\textwidth]{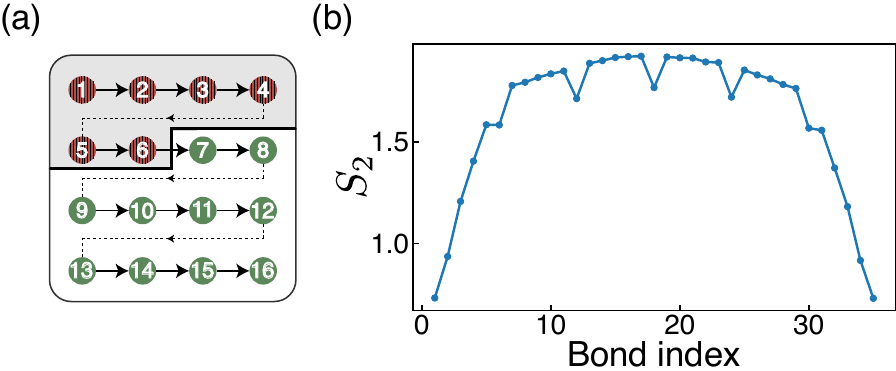}
    \caption{(a) Schematic showing the raster scan ordering and one possible bipartition for conditional sampling. (b) Second R\'{e}nyi entropy $S_2$ for the ground state of the 2D AFH model with system size $6\times 6$. The network is trained with raster scan ordering and the entropy is calculated with conditional sampling.}
    \label{fig:reshape}
\end{figure}

\subsection{Input spin ordering}
To represent a many-body state using NAQS, there are different possible choices of input spin ordering, corresponding to different decompositions of the state distribution into conditional probabilities. Specific ordering choices are only required for conditional sampling, not direct sampling. To perform conditional sampling $\sigma_b \sim p(\sigma_b|\sigma_a)$, only bipartitions along the input spin ordering of the autoregressive network are possible. For direct sampling it is possible to specify arbitrary groups of spins as region $A$ since we only sample from the joint state distribution rather than its conditionals.  

In one dimension, there exists a natural ordering which follows the actual lattice structure and preserves the locality of the interactions. In two dimensions, a raster scan ordering from the top-left to bottom-right [Fig.~\ref{fig:reshape}(a)] seems natural. However, that only allows CS entropy calculations for regions with approximately constant boundary length. The results for $S_2$ calculated from a NAQS trained with the raster scan ordering clearly show features of the limited boundary length [Fig.~\ref{fig:reshape}(b)]. To estimate entropy for square regions of increasing area, we therefore choose the spiral ordering for the 2D lattice as shown in Fig.~4(a) of the main text.

\section{Statistical error bars}
We calculate the variance of both conditional and direct sampling estimators from a bootstrapping analysis. The error bars for conditional sampling results in Fig.~\ref{fig2}, Fig.~\ref{fig3}(c), Fig.~\ref{fig4}(b) and (d) and for low-order direct sampling results in Fig.~\ref{fig2}, Fig.~\ref{fig3}(a-b) are smaller than the marker size and therefore not shown. The statistical error bars are shown for conditional sampling results in Fig.~\ref{fig4}(c) and (e). At large $n$, direct sampling fails to converge because drawing samples that contribute significantly to the estimator mean becomes exponentially less likely. As a result, the error bars cannot be faithfully estimated with the sample sizes we have access to and are therefore also not shown in Fig.~\ref{fig2} and Fig.~\ref{fig4}(c).

\bibliography{references.bib,SI.bib}

\end{document}